\documentclass[twocolumn,showpacs,preprintnumbers,prb]{revtex4}
\usepackage{graphicx}
\usepackage{dcolumn}
\usepackage{bm}
\usepackage{amsmath}
\usepackage{color}
\newcommand{\cdcro}{CdCr$_2$O$_4$}
\newcommand{\zncro}{ZnCr$_2$O$_4$}

\begin{document}
\preprint{}
\title{Synchrotron x-ray study of lattice vibrations in \cdcro}
\author{J.-H. Kim$^{1,2}$}
\author{M. Matsuda$^3$}
\author{H. Ueda$^4$}
\author{Y. Ueda$^4$}
\author{J.-H. Chung$^5$}
\author{S. Tsutsui$^6$}
\author{A. Baron$^{7,8}$}
\author{S.-H. Lee$^1$}
\affiliation{$^1$Department of Physics, University of Virginia, Charlottesville, VA 22904-4714, USA\\
$^2$Max-Planck-Institute for Solid State Research, Stuttgart, 70569, Germany\\
$^3$Quantum Beam Science Directorate, Japan Atomic Energy Agency, Tokai, Ibaraki 319-1195, Japan\\
$^4$Institute for Solid State Physics, University of Tokyo, Kashiwa, Chiba 277-8581, Japan\\
$^5$Department of Physics, Korea University, Seoul 136-713, Korea\\
$^6$SPring-8/JASRI 1-1-1 Kouto, Sayo-cho, Sayo-gun, Hyogo 679-5198, Japan\\
$^7$SPring-8/RIKEN 1-1-1 Kouto, Sayo-cho, Sayo-gun, Hyogo 679-5148, Japan}
\date{\today}

\begin{abstract}
Using inelastic x-ray scattering we have investigated lattice vibrations in a geometric frustrated system \cdcro~that upon cooling undergoes a spin-Peierls phase transition at T$_\text{N}$ = 7.8 K from a cubic and paramagnetic to a tetragonal and Neel state. Phonon modes measured around Brillouin zone boundaries show energy shifts when the transition occurs. Our analysis shows that the shifting can be understood as the ordinary effects of the lowering of the crystal symmetry.
\end{abstract}

\maketitle

A spinel AB$_2$O$_4$ system is an excellent model to study the physics of frustration. The octahedral B sites surrounded by oxygen ions form a three dimensional network of corner sharing tetrahedra, called a pyrochlore lattice. Since oxygen octahedra form an edge sharing network, if the B site is occupied by a magnetic ion with unpaired t$_{2g}$ electrons, the nearest neighbor interactions become dominant, which yields strong frustration. In spite of the theoretical predictions that the pyrochlore system should not order at any temperature, spinels usually undergo phase transitions into ordered states at nonzero temperatures~\cite{Lee09}. This is because the spin degree of freedom can be coupled with other degrees of freedom, such as orbital and lattice, to lift the magnetic degeneracy~\cite{Balents10}. 

One example is a spin-Peiels phase transition that occurs in ACr$_2$O$_4$. The Cr-based spinels in the absence of the orbital degree of freedom, ACr$_2$O$_4$, remain paramagnetic to temperatures far below the Curie-Weiss temperature, -390 K and -88 K for A = Zn~\cite{Lee00} and Cd~\cite{Chung05}, respectively. Upon further cooling the system undergoes a first order spin-Peierls-like phase transition from a cubic paramagnet to a tetragonal Ne\'{e}l state at T$_\text{N}$ = 12.5 K and 7.8 K for A = Zn~\cite{Lee00} and Cd~\cite{Chung05}. The tetragonal lattice distortion induces an exchange anisotropy that lifts the frustration and allows the system to select a particular spin configuration as its ground state. The nature of the phase transition and that of the ground state depend on the delicate balancing act between the lattice energy cost for the distortion and the magnetic energy gain due to the spin ordering. Previous studies have shown that the ionic size of the A ion has a crucial role in the selection process. In the case of \zncro~with smaller Zn$^{2+}$~ions, the phase transition involves a tetragonal contraction along the c-axis with $I\bar{4}m2$~symmetry~\cite{Ji09} and commensurate Ne\'{e}l ordering~\cite{Lee00}. In the case of \cdcro~with larger Cd$^{2+}$~ions, however, the transition yields a tetragonal elongation along the c-axis with  $I4_1 / amd$~symmetry and an incommensurate Ne\'{e}l ordering~\cite{Chung05}. Their microscopic mechanisms are not to be understood yet. A theory based on Dzyaloshinskii-Moriya interactions was proposed to explain the {\it static} spin-lattice coupling in \cdcro~\cite{Chern06}. This theory, however, does not provide an accurate account of the one-to-one correspondence between the tetragonal distortion and the Ne\'{e}l state that were experimentally observed, and the microscopic mechanism of the {\it static} spin-lattice coupling is yet to be understood. 

\begin{figure}
\begin{center}
\includegraphics[width=\hsize]{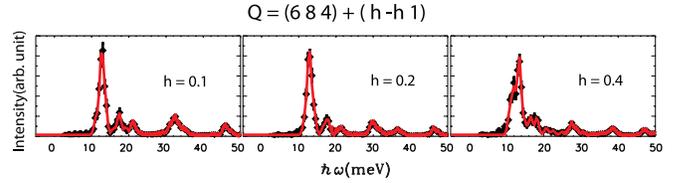}
\caption{(Color online) Phonon spectra measured at 10 K from a single crystal of \cdcro~along $\vec{Q}$=(6,8,4)+(h,-h,1) direction. Each solid (red) line shown is a fit to Gaussians.}
\label{phonon_raw}
\end{center}
\end{figure}

Another issue is whether or not the spin-lattice coupling is also dynamic in nature. Recently, reflectivity data obtained from samples of \zncro~\cite{Sushkov05, Rudolf07, Kant09} and \cdcro~\cite{Sushkov08, Rudolf07, Kant09} have been presented as an evidence for the spin-phonon coupling in these materials. For \cdcro, one of the modes around 45 meV splits into two modes while the other three modes do not, below T$_\text{N}$. The 45 meV mode is a F$_{1u}$ symmetry mode that dominantly involves displacements of the magnetic Cr$^{3+}$ ions. Throughout comparing the intensities of the two split modes, they concluded that the higher energy mode of the two is doublet and the lower energy mode is singlet. The $c$-axis elongation in the tetragonal phase of \cdcro, however, should yield a lower energy doublet mode (yz and zx) and a higher energy singlet mode (xy). This led them to conclude that the phonon anomalies could not be simply explained by the tetragonal distortion rather than spin-phonon coupling must be involved~\cite{Sushkov08}. However, the reflectivity spectra is a limited technique in studying the lattice dynamics because it is a probe of phonons at a zone center ($\vec{Q}$ = 0). 

Here we report our inelastic synchrotron x-ray scattering (IXS) measurements on a single crystal of \cdcro. We characterized the lattice vibrations of  \cdcro~both in the cubic $Fd\bar{3}m$~and the tetragonal $I4_1 / amd$~phases by mapping out the phonon dispersion relations up to 50 meV along high symmetry directions. Our analysis of the data done using a computer software, GULP~\cite{Gale92}, leads us to conclude that the energy shifting of the phonon modes around a zone boundary is simply due to the ordinary effects of the lowering of the crystal symmetry.

A $\sim$ 100 mg single crystal of \cdcro~(space group $Fd\bar{3}m$, a=8.58882 $\mathring{A}$ for T=10 K) was grown using a flux method~\cite{Ueda05, Dabkowska81}. The IXS measurements were done at the high resolution BL35XU beamline at the SPring-8 in Japan. Using the backscattering geometry allows obtaining a large angular acceptance for both the monochromator and the analyzer and an excellent energy resolution of $\sim$ 1.6 meV over the entire range of energy up to $\sim$ 50 meV. A Si(11 11 11) monochromator was set to produce incident x-ray with $E_i= 21.747$ keV onto the sample.

\begin{figure}
\includegraphics[width=\hsize]{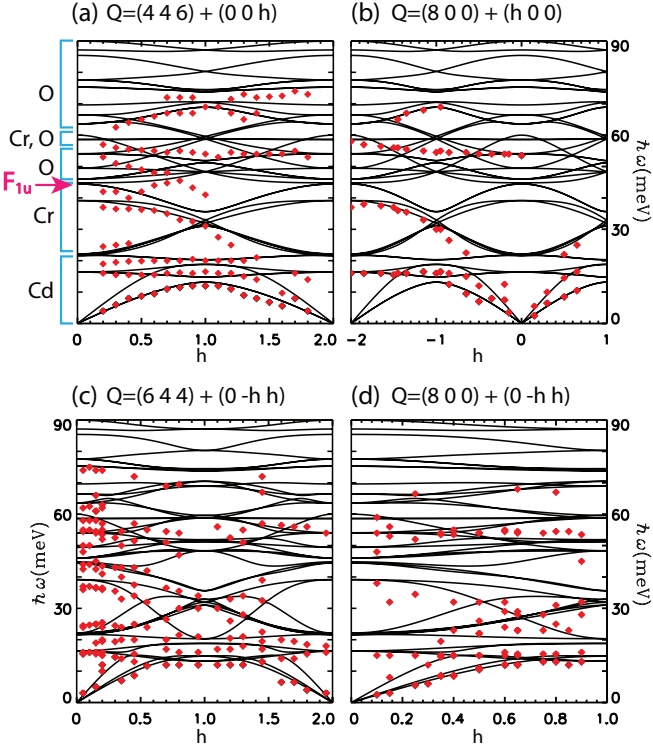}
\caption{(Color online) Phonon dispersion relations as a function of $\vec{Q}$ and $\hbar\omega$ obtained at 10 K \cdcro. Red diamonds indicate the experimental values while black lines show calculated dispersions. (a) Along $\vec{Q}$=(4, 4, h), (b) along $\vec{Q}$=(h, 0, 0), (c) along $\vec{Q}$=(6, 4-h, 4+h) and (d) along $\vec{Q}$=(8, -h, h). Here $\vec{Q}=\vec{\tau}+\vec{q}$ where $\vec{\tau}$ is a reciprocal lattice vector.}
\label{phonon_10}
\end{figure}

We have first characterized the lattice vibrations in the cubic phase. Fig. \ref{phonon_raw} shows the phonon spectra measured at several different wave vectors along the (h,-h,1) direction in the cubic phase. Similar measurements were done along the (h,0,0) and (0,-h,h) directions. A primitive cell of the cubic ($Fd\bar{3}m$) crystal structure has two formula units of \cdcro~and so 14 atoms. Each atom generates three vibrational modes, thus there are 42 phonon normal modes in total. Thus, we have fit the data with 42 gaussians which correspond to 42 possible phonon modes with a fixed energy resolution. The results of the fits are summarized in Fig. \ref{phonon_10} as a function of momentum and energy. 

\begin{figure}
\includegraphics[width=\hsize]{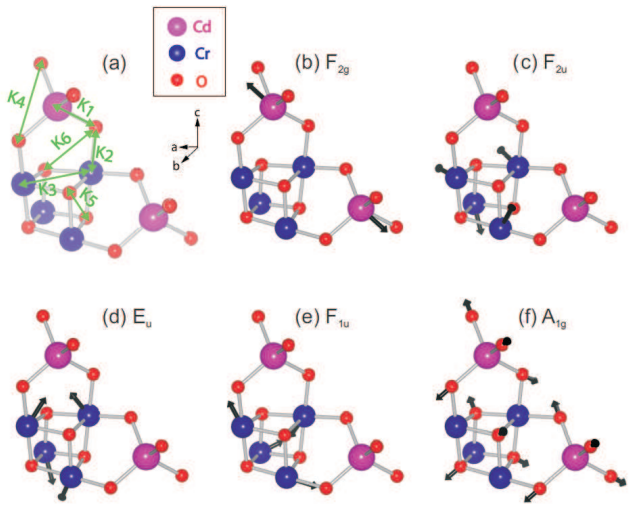}
\caption{(Color online) Six short-range force constants used for a rigid ion model phonon calculation described in the text (a) and the atom vibrations in \cdcro~phonon modes at  $\Gamma$ point  (b-f) in the cubic phase. (b) Cd vibrations dominant mode of F$_{2g}$ at 16.3 meV. Cr vibrations in (c) F$_{2u}$ at 22.1 meV, (d) E$_{u}$ at 39.1 meV, and (e) F$_{1u}$ mode at 44 meV. (f) O vibrations in A$_{1g}$ mode at 87.1 meV. Black arrows are directions of the vibrations.}
\label{cdcro_disp}
\end{figure}

The dispersion relations of the phonon modes have been analyzed using group theory. Theoretical group calculations for $Fd\bar{3}m$ crystal structure yield 39 optical modes at a zone center ($\Gamma$ point). The optical modes are made of the following symmetries,
\begin{align}
\Gamma=  A_{1g}+E_{g}+F_{1g}+3F_{2g} +2A_{2u}+2E_{u}+4F_{1u}+2F_{2u},
\label{eq:prep10}
\end{align}
where the symbols, A, E, and F, represent singlet, doublet, and triplet, respectively. Among them, the $A_{1g}, E_{g}$, and $F_{2g}$ modes are Raman active, the $F_{1u}$ modes are infrared active, and the rest are inactive in both Raman and infrared reflectivity measurements. The dispersions of the phonon modes were calculated by the computer software package, GULP~\cite{Gale92}, with a rigid ion potential model. In this model, the potential energy for the lattice vibrations can be written as 
\begin{align}
V=  \sum_{ l, l'}  \left \{ \frac{1}{2} K_{ll'} (\vec{r}_{ll'}-\vec{r}_{0_{ll'}})^2 -\frac{Z_l Z_{l'}}{r_{ll'}} \right \},
\label{eq:cdcro_potential}
\end{align}
where K$_{ll'}$ is a short range force constant, ($\vec{r}_{ll'}-\vec{r}_{0_{ll'}}$) is the displacement vector from the equilibrium position of the $l$-th and $l'$-th atoms and Z$_l$ is an effective charge on the $l$-th atom. The first term of eq. (\ref{eq:cdcro_potential}) corresponds to the short range repulsive forces, while the second part is due to the long range Coulomb interactions. For simplicity, the Coulomb interactions are considered only between the nearest neighbor atoms. In the calculations we used six short-range force constants described in Fig. \ref{cdcro_disp} (a). An initial set of those constants was used to generate a reasonable fit to the experimental data, and the experimental data were re-fitted to 42 gaussians to obtain the peak positions, i.e., characteristic energies of the phonon modes at different Qs. Because some of the modes are too close in energy to be distinguished by the experimental resolution, the number of peaks was reduced to 16 or 20 depending on which Q when the data were refitted to 42 Gaussian. We repeated this process several times until the comparison between the calculated dispersions and the experimental data were converged. The final optimal values of the force constants are listed in Table \ref{cdcro_para} and the resultant dispersion relations are shown by the black lines in Fig. \ref{phonon_10}.

\begin{table}
\begin{center}
\caption{The values of the short-range force constants and effective dynamical ionic charges}
\begin{tabular}{cccc}
Force constants & Values   & Effective charges & Values \\ 
  &   (eV\AA) &   &  (e)\\ \hline
Cd-O (K$_1$) & 11.22 & Cd & 0.44\\
Cr-O (K$_2$) & 9.77 & Cr & 1.64\\
Cr-Cr (K$_3$) & 0.98 & O & -0.93\\
O-O (K$_4$) & 0.46   \\
O-O (K$_5$) &  0.70\\
O-O (K$_6$) &  0.00 \\
\hline
\end{tabular}
\label{cdcro_para}
\end{center}
\end{table}

\begin{table}
\begin{center}
\caption{Calculated phonon energies at a zone center above and below T$_\text{N}$ in \cdcro. When the system transits from cubic to tetragonal, most phonon modes change only by a few tenth percent of their energies, consistent with the few tenth percent change in the atomic distances ($\hbar \omega \propto r^{-3/2}$). The oxygen-stretching A$_{1g}$~and A$_{2u}$~modes, however, change by $\sim$ 2 \% in energy. This can be understood by the fact that the atomic distances between oxygen atoms reduce by 0.8 \% (e.g. 2.596 \AA~in cubic to 2.576 \AA~in tetragonal phase) so that the energy changes by 1.3 \%}
\begin{tabular}{cc|cccccccc}
 \hline  
\multicolumn{2}{c|}{$Fd\bar{3}m$, 10 K} & \multicolumn{2}{c}{$I4_1 / amd$, 6 K} \\ \hline
Species & E (meV) & E (meV) & Species \\  \hline
 A$_{1g}$  & 87.1& 85.3 & A$_{1g}$    \\ \hline
A$_{2u}$ & 60.0    & 60.0   & B$_{1u}$ \\ \hline
              &  89.9  &   88.1 &              \\ \hline
E$_{g}$ & 49.4 & 48.2, 49.1 & A$_{1g}$ + B$_{1g}$ \\ \hline
E$_{u}$ & 39.1  & 39.1, 39.3  & A$_{1u}$ + B$_{1u}$  \\  \hline
             &   66.4  &   65.6, 66.3&    \\  \hline
F$_{1g}$  & 54.0 & 54.1, 54.2  & A$_{2g}$ + E$_{g}$ \\ \hline
F$_{2g}$ & 16.3  & 16.2, 16.4   &  B$_{2g}$ + E$_{g}$ \\
             &  63.5 &  63.5, 63.7  &   \\ 
		& 77.5 & 76.0, 77.5 & \\ \hline
F$_{1u}$ (TO)  & 21.6 & 21.1,21.2 & A$_{2u}$ + E$_{u}$ (TO) \\ 
		&   44.4 &   44.3, 44.4&\\ 
		& 58.6   & 58.3, 58.7   & \\ 
		& 75.2  &  74.2, 75.1 &\\ \hline
F$_{1u}$ (LO) 	& 21.7   & 21.7  & A$_{2u}$ + E$_{u}$ (LO)\\ 
		&  44.8 &  44.6 &  \\
		& 69.6 & 68.6 &  \\
		&  85.3 &  85.3& \\  \hline
F$_{2u}$  & 22.1   & 22.3, 22.3  & B$_{2u}$ + E$_{u}$ \\  
		&  46.0 & 46.0, 46.1 &\\  \hline
\end{tabular}
\label{cdcro_group}
\end{center}
\end{table}


\begin{figure}
\includegraphics[width=0.9\hsize]{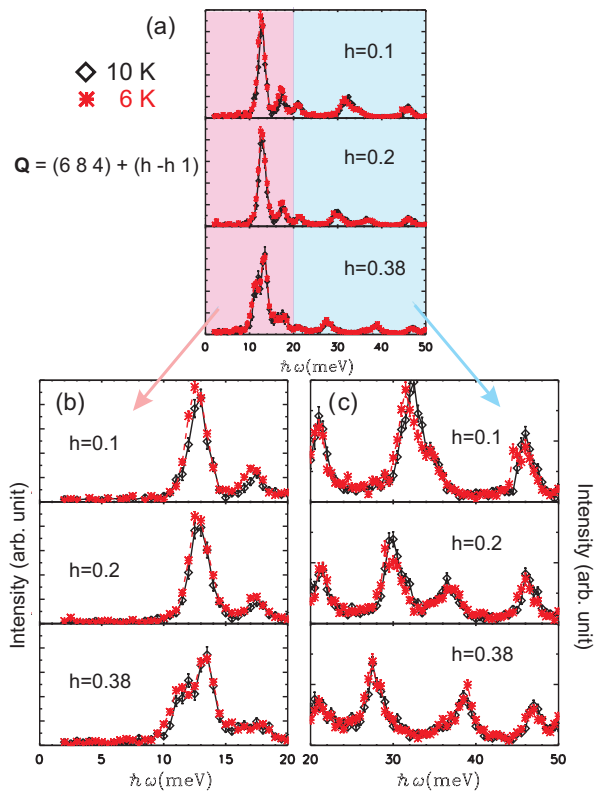}
\caption{(Color online) The IXS phonon spectra as a function of energy, $\hbar\omega$, obtained from a single crystal \cdcro~along $\vec{Q}$=(6,8,4)+(h,-h,1) at 10 K (black diamond) and 6 K (red asterisk), in the energy range (a) from 2 to 50 meV, (b) from 2 to 20 meV: Cd dominant modes, and (c) from 20 to 50 meV : Cr dominant modes. }
\label{phonon_6_exp}
\end{figure}

Table \ref{cdcro_group} lists the energies of the phonon modes at a $\Gamma$ point. The values of energies are consistent with the previous phonon studies on other spinel compounds~\cite{Zwinscher95,Lutz90}. Here, at a $\Gamma$ point the ($F_{2u}~\text{and}~F_{1u}$) modes with $0 < \hbar\omega < 22$ meV involve mainly vibrations of the heavy Cd ions, the ($F_{1u}$, $F_{2g}$, and $E_{u}$) modes with $22 < \hbar\omega < 45$ meV involve mainly vibrations of the Cr ions, while the modes with $\hbar \omega > 45$ meV involve mainly O ions except modes with  $\hbar \omega \sim 60$ meV which have dominant Cr vibrations. Thus, the ($F_{1u}$, $F_{2g}$, and $E_{u}$) modes are expected to show anomalous behaviors when the system undergoes the phase transition if spin-phonon coupling is involved. Eigenstates of some typical vibrational modes at the zone center are illustrated in Fig. \ref{cdcro_disp}. 

\begin{figure}
\includegraphics[width=0.9\hsize]{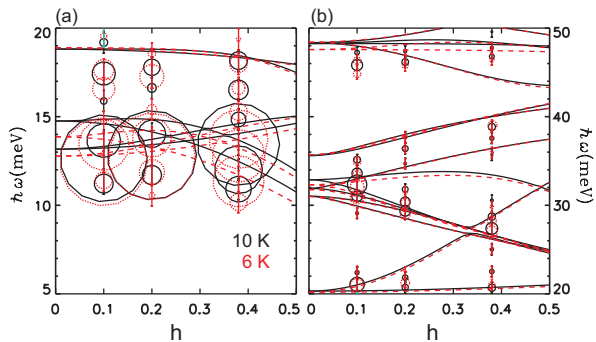}
\caption{(Color online) The peak positions of phonon modes (circles) and calculated dispersions as a function of  $\vec{q}$=(h,-h,1) at 10 K (black solid lines) and 6 K (red dashed lines). The size of each circle is proportional to the integrated intensity of each peak.  (a) and (b) show the dominantly Cd and Cr ion vibrational modes respectively. }
\label{phonon_6_disp}
\end{figure}

After the careful characterization of the lattice vibrations in the cubic phase, we have also studied the tetragonal phase. First, we confirmed the symmetry of the tetragonal structure to be $I4_1/amd$ not $I4_122$ by finding out that elastic nuclear reflections are absent at $(6,0,0)_\text{cubic} = (3,3,0)_\text{tetra}$ and $(8,2,0)_\text{cubic} = (3,5,0)_\text{tetra}$. Second, we have performed IXS measurements to study lattice vibrations in the tetragonal phase. Fig. \ref{phonon_6_exp} (a) shows the data obtained upto 50 meV at 6 K ($<$ T$_\text{N}$), over-plotted with the 10 K data of the cubic phase at three different wave vector positions along $\vec{q}$ = (h, -h, 1). As shown in Fig. \ref{phonon_6_exp}  (b) and \ref{phonon_6_disp} (a), among the mainly Cd vibrating modes upto 20 meV, two modes show changes between the two temperature data: for $\vec{Q}$ = (6.1, 7.9, 5) the peak centered around 18 meV becomes broad at 6 K and for $\vec{Q}$ = (6.38, 7.62, 5) the peak centered at 12.0 meV at 10 K shifts to 11.4 meV at 6 K. In Fig. \ref{phonon_6_exp} (c) and \ref{phonon_6_disp} (b) which show the mainly Cr vibrating modes, the peak that appears at 32.1 meV at 10 K moves to 31.0 meV at 6 K for $\vec{Q}$ = (6.1, 7.9, 5). At the same $\vec{Q}$ position, the two peaks that appear at 45.8 and 47.2 meV  at 10 K move to 44.8 and 46.3 meV below T$_\text{N}$. The other peaks do not show any obvious difference between two temperatures. Similar measurements were also performed along $\vec{Q}$ = (7,7,5)+(h,h,0.2), and only those four modes seem to shift down by $\sim$1 meV when the temperature changes from 10 K to 6 K. Now, the question is that the observed changes in phonon spectra are simply either due to the cubic-to-tetragonal structural distortion or a spin-phonon coupling. In order to address this issue, we used the rigid ion model to reproduce the phonon spectra in the tetragonal phase.

There are 39 optical modes in the $I4_1/amd$~tetragonal phase at a zone center, which can be described as
\begin{align}
\Gamma = & 2A_{1g}+2A_{1u}+A_{2g}+4A_{2u} \nonumber \\
& +B_{1g}+4B_{1u} +3B_{2g}+2B_{2u}+4E_{g}+6E_{u}.
\label{eq:prep6}
\end{align}
Therefore, the triply-degenerate modes split into a singlet and a doublet mode in the tetragonal phase, and the doubly-degenerate modes split into two singlet modes. In the cubic phase, the distance between nearest neighbor Cr ions is $r_{NN} = 3.037$\AA. In the tetragonal phase, $r_{NN} =  3.034 $\AA~is for the bonds in the $ab$-plane and $r_{NN} = 3.040$\AA~is for the bonds in the $ac$- and $bc$-planes. Since the difference in $r_{NN}$ is small, we assumed that the force constants, K$_{ll'}$, remain unchanged after the tetragonal distortion. The effect of the lattice distortion is reflected as the changes in Coulomb forces. The calculated energies of the phonon modes at the $\Gamma$ point for the tetragonal phase are listed in Table \ref{cdcro_group}, and their dispersions are shown by dashed (red) lines in Fig. \ref{phonon_6_disp}. Even though the simple rigid ion model does not perfectly reproduce our data, the calculation shows that the four modes around 13.2, 14.8, 33 and 48 meV at $\vec{q}=(0.1,-0.1,1)$~position in cubic phase shift to 12.8, 14, 32.5 and 47 meV in tetragonal phase. This is consistent with our experimental observation at the same $\vec{q}$ point where the modes with $\hbar \omega \sim$ 13, 32, and 46 meV  at 10 K shift to lower energies by $\sim$ 1 meV at 6 K. The observed changes in our x-ray phonon spectra seem to be due to the ordinary mode shifting by the lowering of the crystal symmetry, rather than the spin-phonon coupling, which was previously claimed by reflectivity measurements~\cite{Sushkov08, Rudolf07, Kant09}. It remains to be seen whether or not a zone-center and a zone-boundary should behave differently in this compound.

In conclusion, we have searched possible phonon anomalies that might be associated with the origin of the phase transitions in the frustrated magnet \cdcro~using inelastic synchrotron x-ray scattering. This was to address the issue of whether or not a static as well as dynamic spin-lattice coupling is part of the mechanism leading to the phase transition. Our results show that a phonon mode energy around a zone boundary with $\hbar\omega \sim$ 45 meV shifts down as the crystal structure changes from cubic to tetragonal. The shifting can be well reproduced by a rigid ion model with the same force constants as used for the cubic phase.

Work at the University of Virginia is supported by the U.S. Department of Energy, Office of Basic Energy Sciences, Division of Materials Sciences and Engineering, under Contract No. DE-FG02-07ER46384. JHC is supported by the Nuclear R\&D Program (No. 2010-0018369) of NRF Korea grant funded by the Korea government (MEST).

\end{document}